\documentclass{webofc}
\usepackage[varg]{txfonts}   
\begin{document}
\title{SUSY-like relation of the splitting functions in evolution of gluon and quark jet multiplicities}
%
%

\author{\firstname{Anatoly} \lastname{Kotikov}\inst{}\fnsep\thanks{\email{kotikov@theor.jinr.ru}}
}

\institute{
  Laboratory of Theoretical Physics,
Joint Institute for Nuclear Research, 141980 Dubna, Russia
          }

\abstract{%
We show
the new
relationship \cite{Kniehl:2017fix} between the anomalous
dimensions, resummed through next-to-next-to-leading-logarithmic order, in the
Dokshitzer-Gribov-Lipatov-Altarelli-Parisi (DGLAP) evolution equations for the
first Mellin moments $D_{q,g}(\mu^2)$ of the quark and gluon fragmentation
functions, which correspond to the average hadron multiplicities in jets
initiated by quarks and gluons, respectively.
So far, such relationships have only been known from supersymmetric (SUSY) QCD.
Exploiting available next-to-next-to-next-to-leading-order (NNNLO) information on
the ratio $D_g^+(\mu^2)/D_q^+(\mu^2)$ of the dominant plus components,
the  fit of the
world data of $D_{q,g}(\mu^2)$ for charged hadrons measured in $e^+e^-$
annihilation leads to
$\alpha_s^{(5)}(M_Z)=0.1205\genfrac{}{}{0pt}{}{+0.0016}{-0.0020}$. 
}
\maketitle

In
QCD \cite{Bjorken:1969ja}, the inclusive
production of single hadrons involves the notion of fragmentation functions
$D_a(x,\mu^2)$, where $\mu$ is the factorization scale.
Owing to the factorization theorem, the $D_a(x,\mu^2)$ functions are universal
in the sense that they do not depend on the process by which parton $a$ is
produced.
By local parton-hadron duality \cite{Azimov:1984np}, there should be a local
correspondence between parton and hadron distributions in hard-scattering
processes.
So,
$D_a(x,\mu^2)$ are genuinely nonperturbative
and need to be determined by fitting
experimental data.
However, once $D_a(x,\mu_0^2)$ are assumed to be known, their $\mu^2$
dependences
are governed by the timelike DGLAP
evolution equations \cite{Gribov:1972ri,Dokshitzer:1977sg}, whose
splitting functions $P_{ba}(x)$
are known at next-to-next-to-leading order \cite{Almasy:2011eq}.
The scaling violations, {\it i.e.}, the $\mu^2$ dependences, of $D_a(x,\mu^2)$
may be exploited in global data fits to extract the strong-coupling constant
$\alpha_s=g_s^2/(4\pi)$, leading to very competitive results
\cite{Kniehl:2000fe} as for the world average \cite{Olive:2016xmw}.

The DGLAP equations are conveniently solved in Mellin space, where
$D_a(N,\mu^2)=\int dx\,x^{N-1}D_a(x,\mu^2)$ with $N=1,2,\ldots$ and similarly for
$P_{ba}(x)$:
\begin{equation}
\frac{\mu^2d}{d\mu^2}
\left(\begin{array}{l} D_s(N,\mu^2) \\ D_g(N,\mu^2) \end{array}\right)
=\left(\begin{array}{ll} P_{qq}(N) & P_{gq}(N) \\
P_{qg}(N) & P_{gg}(N) \end{array}\right)
\left(\begin{array}{l} D_s(N,\mu^2) \\ D_g(N,\mu^2) \end{array}\right),
\label{apR}
\end{equation}
where $D_s=(1/2n_f)\sum_{q=1}^{n_f}(D_q+D_{\bar{q}})$, with $n_f$ being the
number of active quark flavors, is the quark singlet component.
The quark non-singlet component
is irrelevant for the following.
After solving the DGLAP equations in Mellin space, one returns to $x$ space via
the inverse Mellin transform, analytically continuing $N$ to complex values.

The first Mellin moment $D_a(\mu^2)\equiv D_a(1,\mu^2)$ is of particular
interest in its own right because
it corresponds to the average hadron multiplicity
$\langle n_h\rangle_a$ of jets initiated by parton $a$.
There exists a wealth of experimental data on $\langle n_h\rangle_q$,
$\langle n_h\rangle_g$, and their ratio
$r=\langle n_h\rangle_g/\langle n_h\rangle_q$ for charged hadrons $h$ taken in
$e^+e^-$ annihilation at various center-of-mass energies $\sqrt{s}$, ranging
from 10 to 209~GeV (for a comprehensive compilation of experimental
publications, see Ref.~\cite{Bolzoni:2013rsa}), which allows for a
high-precision determination of $\alpha_s$
\cite{Bolzoni:2013rsa,Perez-Ramos:2013eba}.
This provides a strong motivation for us to deepen our theoretical
understanding of $D_a$ within the QCD formalism as much as possible, which is
actually limiting the error in the value of $\alpha_s$ thus extracted.
The study of $D_a$ is a topic of old vintage; the LO value of $r$,
$C^{-1}=C_A/C_F$ with color factors $C_F=4/3$ and $C_A=3$, was found four
decades ago \cite{Brodsky:1976mg}.
Subsequent analyses \cite{Perez-Ramos:2013eba,Malaza:1985jd} were performed
using the generating-functional approach in the modified leading-logarithmic
approximation (MLLA) \cite{Dokshitzer:1991wu}.

The description of the $\mu^2$ dependences of $D_a$ at fixed order in
perturbation theory are spoiled by the fact that $P_{ba}\equiv P_{ba}(1)$ are
ill defined and require resummation, which was performed for the leading
logarithms (LL) \cite{Mueller:1981ex}, the next-to-leading logarithms (NLL)
\cite{Vogt:2011jv}, and the next-to-next-to-leading logarithms (NNLL)
\cite{Kom:2012hd}.
In Refs.~\cite{Bolzoni:2013rsa,Bolzoni:2012ii}, Eq.~(\ref{apR}) is first
diagonalized for arbitrary value of $N$ at LO, and then the NNLL resummation is
incorporated.
Unfortunately, this two-step procedure, which has been standard practice in the
literature so far \cite{Buras:1979yt,Ellis:1993rb}, fails to fully exploit the
available knowledge on the higher-order corrections and yields an
approximation, the uncertainty of which is difficult to estimate reliably.

In Ref. \cite{Kniehl:2017fix} (see also \cite{Kniehl:2017oat}),
we exposed a relationship between the NNLL-resummed expressions
for $P_{ba}$, which has gone unnoticed so far.
Its existence in QCD is quite remarkable and interesting in its own right,
because a similar relationship is familiar from SUSY
QCD,
where $C=1$
\cite{Dokshitzer:1977sg,Dokshitzer:1991wu,Kom:2012hd,Kounnas:1982de}.

Our starting point is Eq.~(\ref{apR}) for $N=1$ with NNLL resummation.
We have \cite{Kom:2012hd} $(a=q,g)$
\begin{eqnarray}
P_{aa} = \gamma_0(\delta_{ag} + K_{a}^{(1)} \gamma_0 
+ K_{a}^{(2)}  \gamma_0^2),~~
P_{gq} = C (P_{gg} +A) + \mathcal{O}(\gamma_0^4),~~
P_{qg} =  C^{-1} (P_{qq} +A),
\label{NNLL}
\end{eqnarray}
with $\mathcal{O}(\gamma_0^3)$ accuracy,
where $\gamma_0=\sqrt{2C_Aa_s}$, with $a_s=\alpha_s/(4\pi)$ being the couplant,
$\delta_{ab}$ is the Kronecker symbol, and $(\varphi= n_f/C_A)$
\begin{eqnarray}
K_{q}^{(1)} &=& \frac{2}{3} C\varphi,~~K_{g}^{(1)} = -\frac{1}{12}[11 +2\varphi (1+6C)],~~
K_{q}^{(2)} = -\frac{1}{6} C\varphi [17-2\varphi(1-2C)],
\nonumber \\
K_{g}^{(2)} &=& \frac{1193}{288} -2\zeta(2)
- \frac{5\varphi}{72}(7-38C)+\frac{\varphi^2}{72}(1-2C)(1-18C),~~
A =  K_{q}^{(1)}\gamma_0^2.
\label{nllfirstA}
\end{eqnarray}
Eq.~(\ref{NNLL}) is written in a form that allows us to glean a novel
relationship:
\begin{equation}
C^{-1}P_{gq}-P_{gg}=CP_{qg}-P_{qq},
\label{Basic}
\end{equation}
which is independent of $n_f$.
Eq.~(\ref{Basic}) generalizes the case of SUSY QCD
\cite{Dokshitzer:1977sg,Dokshitzer:1991wu,Kom:2012hd,Kounnas:1982de} from
$C=1$ to arbitrary $C$ values.

The corresponding relation in $\mathcal{N}=1$ SUSY
\cite{Dokshitzer:1977sg} is known to be violated beyond LO 
\cite{Almasy:2011eq}.
It will be interesting to see if Eq.~(\ref{Basic}) also holds beyond
$\mathcal{O}(\gamma_0^3)$,
et least in the case of the schemes, which preserve supersymmetry
properties, such as the dimensional reduction.
The choice of a scheme in above consideration is not
  so important because a difference in the results of various schemes
  is exactly canceled in Eq. (\ref{Basic}).

  Perhaps, the result (\ref{Basic}) may be relate with Lipatov observation 
\cite{Lipatov:2004pk}
on an integrability in the high-energy limit of QCD. Of course, the
Lipatov observation is based on the resummation of the large $\ln (1/x)$
terms in the space-like kinematics. Here we have a similar resummation
in the time-like kinematics and the possible relation is not so
obvious and should need strong investigations.

We
solve Eq.~(\ref{apR}) for $N=1$ exactly by exploiting Eq.~(\ref{Basic}).
To this end, we diagonalize the NNLL DGLAP evolution kernel as
\begin{equation}
U^{-1}\left(\begin{array}{ll}
P_{qq} & P_{gq} \\ P_{qg} & P_{gg}
\end{array}\right)U
=\left(\begin{array}{ll}
P_{--} & 0 \\ 0 & P_{++}
\end{array}\right),~~
U=\left(\begin{array}{ll}
1 & -1 \\ \frac{1-\alpha}{\varepsilon} & \frac{\alpha}{\varepsilon}
\end{array}\right),~~
U^{-1}=\left(\begin{array}{ll} 
\alpha & \varepsilon \\ \alpha-1 & \varepsilon
\end{array}\right),
\label{matrix}
\end{equation}
where 
\begin{equation}
\alpha =\frac{P_{qq}-P_{++}}{P_{--}-P_{++}},~~
\varepsilon=\frac{P_{gq}}{P_{--}-P_{++}},~~
P_{\pm\pm} = \frac{1}{2}\left[P_{qq}+P_{gg}\pm
\sqrt{(P_{qq}-P_{gg})^2+4P_{qg}P_{gq}}\right].\qquad
\label{Ppm}
\end{equation}
Eq.~(\ref{apR}) for $N=1$ thus assumes the form
\begin{equation}
\frac{\mu^2d}{d\mu^2}\left(\begin{array}{l} D_- \\ D_+ \end{array}\right)
=\left[
\left(\begin{array}{ll} P_{--} & 0 \\ 0 & P_{++}\end{array}\right)
-U^{-1}\frac{\mu^2d}{d\mu^2}U\right]
\left(\begin{array}{l} D_- \\ D_+ \end{array}\right),
\label{ap2a}
\end{equation}
where the second term contained within the square brackets stems from the
commutator of $\mu^2d/d\mu^2$ and $U$, and
\begin{equation}
\left(\begin{array}{l} D_- \\ D_+ \end{array}\right)
=U^{-1}\left(\begin{array}{l} D_s \\ D_g \end{array}\right)
=\left(\begin{array}{l}(\alpha D_s+\varepsilon D_g \\
\alpha-1)D_s+\varepsilon D_g \end{array}\right).
\label{ap1.2}
\end{equation}
Owing to Eq.~(\ref{Basic}), the square root in
Eq.~(\ref{Ppm})
is exactly canceled,
and we have
simple expressions for $P_{\pm \pm}$
\begin{eqnarray}
P_{--} = -A,~~ P_{++}=P_{qq}+P_{gg}+A,~~
\alpha = \frac{P_{gg}+A}{P_{qq}+P_{gg}+2A},~~
\varepsilon = -C \alpha \, .
\label{alpha1}
\end{eqnarray}
Inserting
Eq.~(\ref{alpha1}) in Eq.~(\ref{matrix}), we
have
\begin{equation}
U^{-1}\frac{\mu^2d}{d\mu^2}U=-\frac{1}{\alpha}\,\frac{\mu^2d}{d\mu^2}\alpha
\left(\begin{array}{ll} 1 & 0 \\ 1 & 0\end{array}\right).
\label{AddTerm}
\end{equation}
Using the QCD $\beta$ function,
\begin{equation}
\frac{\mu^2d}{d\mu^2}a_s=\beta(a_s)=-\beta_0a_s^2-\beta_1a_s^3+\mathcal{O}(a_s^4),~~
\beta_0=\frac{C_A}{3}(11-2\varphi),~~
\beta_1=\frac{2C_A^2}{3}[17-\varphi(5+3C)],
\label{eq:beta}
\end{equation}
after a small algebra
we may cast Eq.~(\ref{apR}) in its final form,
\begin{equation}
\frac{\mu^2d}{d\mu^2}
\left(\begin{array}{l} D_- \\ D_+ \end{array}\right)
= \left(\begin{array}{ll} \frac{C\varphi\beta_0}{3C_A}\gamma_0^3-A & 0 \\ 
\frac{C\varphi\beta_0}{3C_A}\gamma_0^3 & P_{gg}+P_{qq} + A \end{array}\right) 
\left(\begin{array}{l} D_- \\ D_+ \end{array}\right).\quad
\label{ap2b}
\end{equation}
The initial conditions are given by Eq.~(\ref{ap1.2}) for $\mu=\mu_0$ in terms
of the three constants $\alpha_s(\mu_0^2)$, $D_s(\mu_0^2)$, and $D_g(\mu_0^2)$.

The solution of Eq.~(\ref{ap2b}) is greatly facilitated by the fact that one
entry of the matrix on its right-hand side is zero.
We may thus obtain $D_-$ as the general solution of a homogeneous differential
equation, 
\begin{equation}
\frac{D_-(\mu^2)}{D_-(\mu_0^2)} =
\exp\!{\left[\int_{\mu_0^2}^{\mu^2}\!\!\!\frac{d\bar{\mu}^2}{\bar{\mu}^2}
\!\left(\frac{C\varphi\beta_0}{3C_A}\gamma_0^3-A\!\right)\!\right]}
= \frac{T_-(\gamma_0(\mu^2))}{T_-(\gamma_0(\mu_0^2))}, 
\label{gensol.-}
\end{equation}
where
\begin{eqnarray}
T_-(\gamma_0) = \exp{\left[\frac{4C\varphi}{3}\int d\gamma_0
\left(\frac{2C_A}{\beta_0\gamma_0 }-1\right)\right]} 
= \gamma_0^{d_-}\exp{\left(-\frac{4}{3}C\varphi \gamma_0\right)},~~ d_-=\frac{8C_AC\varphi}{3\beta_0} \, .
\label{T-}
\label{eq:t-}
\end{eqnarray}
The
correction $\propto\gamma_0$ in Eq.~(\ref{eq:t-}) originates from
the extra term in Eq.~(\ref{ap2a}) and represents a novel feature of our
approach.
In Refs.~\cite{Bolzoni:2013rsa,Bolzoni:2012ii} and analogous analyses for
parton distribution functions \cite{Kotikov:1998qt,Kotikov:2012sm}, the minus
components do not participate in the resummation.

We are then left with an inhomogeneous differential equation for $D_+$.
The general solution $\tilde{D}_+$ of its homogeneous part reads
\begin{eqnarray}
\frac{\tilde{D}_+(\mu^2)}{\tilde{D}_+(\mu_0^2)} =
\exp{\left[\int_{\mu_0^2}^{\mu^2}\frac{d\bar{\mu}^2}{\bar{\mu}^2}
\gamma_0\left(1+ K_+^{(1)}\gamma_0+K_+^{(2)}\gamma_0^2\right)\right]}
= \frac{T_+(\gamma_0(\mu^2))}{T_+(\gamma_0(\mu_0^2))},
\label{gensol.+}
\end{eqnarray}
where
\begin{eqnarray}
K_+^{(1)}&=&2K_{q}^{(1)}+K_{g}^{(1)} =
-\frac{1}{12}[11 +2\varphi (1-2C)],
\nonumber\\
K_+^{(2)}&=&K_{q}^{(2)}+K_{g}^{(2)} 
= \frac{1193}{288} -2\zeta(2) - \frac{7\varphi}{72}(5+2C) 
+\frac{\varphi^2}{72}(1-2C)(1+6C)
\end{eqnarray}
\begin{equation}
T_+(\gamma_0) = \exp{\left[-\frac{4C_A}{\beta_0}
\int\frac{d\gamma_0}{\gamma_0^2}\,
\frac{1+ K^{(1)}_{+}\gamma_0 +  K^{(2)}_{+}\gamma_0^2}{1+b_1\gamma_0^2}
\right]} 
= \gamma_0^{d_+}\exp{\left[\frac{4C_A}{\beta_0\gamma_0}
-\frac{4C_A}{\beta_0} \left(K_+^{(2)}-b_1\right)\gamma_0\right]},
\label{T+}
\end{equation}
with $d_+=-4C_AK_+^{(1)}/\beta_0$ and $b_1=\beta_1/(2C_A \beta_0)$.
Adding to $\tilde{D}_+$ a special solution of the inhomogeneous differential
equation for $D_+$, we find its general solution to be 
\begin{equation}
D_+(\mu^2) = \left[\frac{D_+(\mu_0^2)}{T_+(\gamma_0(\mu_0^2))}
-\frac{4}{3}C\varphi\frac{D_-(\mu_0^2)}{T_-(\gamma_0(\mu_0^2))}
\int_{\gamma_0(\mu_0^2)}^{\gamma_0(\mu^2)}
\frac{d\gamma_0}{1+b_1\gamma_0^2}\,\frac{T_-(\gamma_0)}{T_+(\gamma_0)}\right]
T_+(\gamma_0(\mu^2)).
\label{gensol.+a}
\end{equation}
The final expressions for $D_-$ and $D_+$ in Eqs.~(\ref{gensol.-}) and
(\ref{gensol.+}), respectively, are fully renormalization group improved
because all $\mu$ dependence resides in $\gamma_0$.

Using Eqs.~(\ref{matrix}) and (\ref{ap1.2}), we now return to the parton basis,
where it is useful to decompose $D_a=D_a^++D_a^-$ into the large and small
components $D_a^\pm$ proportional to $D_\pm$, respectively.
Defining $r_\pm=D_g^\pm/D_s^\pm$ and using Eqs.~(\ref{NNLL}), (\ref{nllfirstA}),
and (\ref{alpha1}), we then have $D_s^\pm=\mp D_\pm$ and
\begin{equation}
r_+ = -\frac{\alpha}{\epsilon}=\frac{1}{C}+\mathcal{O}(\gamma_0^2),~~
r_-=\frac{1-\alpha}{\epsilon} =-\frac{4}{3}\varphi\gamma_0
+\frac{\varphi}{18}[29-2\varphi(5-2C)]\gamma_0^2
+\mathcal{O}(\gamma_0^3).
\label{eq:rpm}
\end{equation}
Recalling that $\langle n_h\rangle_q=D_s$ and $\langle n_h\rangle_g=D_g$, we
thus have
\begin{equation}
r= \bigl(r_++r_-D_s^-/D_s^+\bigr)/\bigl(1+D_s^-/D_s^+\bigr).
  \label{eq:r}
\end{equation}
Eq.~(\ref{eq:rpm}) differs from Eqs.~(53) and (54) in
Ref.~\cite{Bolzoni:2013rsa},
\begin{eqnarray}
\bar{r}_+ = \frac{1}{C}\left\{1-\frac{\gamma_0}{3}[1+\varphi(1-2C)]
+\mathcal{O}(\gamma_0^2)\right\},~~
\bar{r}_- = -\frac{2}{3}\varphi\gamma_0+\mathcal{O}(\gamma_0^2).
\label{eq:npb}
\end{eqnarray}
On the other hand, $\bar{r}_+$ in Eq.~(\ref{eq:npb}) agrees with the result for
$r$ obtained in Ref.~\cite{Mueller:1983cq} in the approximation of putting
$D_a^-=0$ and extended to through $\mathcal{O}(\gamma_0^3)$ in
Refs.~\cite{Dremin:1993vq,Capella:1999ms}, which is in line with the reasoning
in Chapter~7 of Ref.~\cite{Dokshitzer:1991wu}.
By the same token, we may accommodate the higher-order corrections
\cite{Dremin:1993vq,Capella:1999ms} by including within the curly brackets in
Eq.~(\ref{eq:npb}) the terms 
$\bar{c}_2\gamma_0^2+\bar{c}_3\gamma_0^3$.
The exact results for $\bar{c}_2$ and
$\bar{c}_3$ can be found in \cite{Kniehl:2017fix}.
For $n_f=5$,
\begin{equation}
\bar{r}_+=2.250-0.889\,\gamma_0-4.593\,\gamma_0^2+0.740\,\gamma_0^3
+\mathcal{O}(\gamma_0^4).
\label{eq:rp5}
\end{equation}
The difference between $r_\pm$ and $\bar{r}_\pm$ is an artifact of the
different
diagonalization procedures adopted here and in Ref.~\cite{Bolzoni:2013rsa}.
In fact, taking the limit $N\to1$ in $D_a(N,\mu^2)$ and diagonalizing the DGLAP
equations are noncommuting operations.
Consequently, our components $D_a$ differ from those in
Ref.~\cite{Bolzoni:2013rsa}, $\overline{D}_a$ with
$\bar{r}_\pm=\overline{D}_g^\pm/\overline{D}_s^\pm$, by terms of
$\mathcal{O}(\gamma_0)$.
Specifically, we have
\begin{eqnarray}
D_a^\pm &=& \sum_{b=s,g}M_{ab}\overline{D}_b^\pm,~~
M_{ss} = 1-\frac{4}{3}C\varphi\gamma_0,~~
M_{sg}= - \frac{C}{3} \gamma_0 [1+\varphi(1-6C)],~~
\nonumber\\
M_{gs} &=& -\frac{2}{3}\varphi\gamma_0,~~
M_{gg}=1+\frac{2}{3}C\varphi\gamma_0.
\label{transform}
\end{eqnarray}
In fact, this transformation converts $\bar{r}_\pm$ into $r_\pm$ and
allows us to extend our result for $r_+$ through
$\mathcal{O}(\gamma_0^3)$; the counterpart of Eq.~(\ref{eq:rp5}) reads 
\begin{equation}
r_+=2.250-4.505\,\gamma_0^2-0.586\,
\gamma_0^3 +\mathcal{O}(\gamma_0^4).
\label{eq:rpcap}
\end{equation}
We denote the approximation of using Eq.~(\ref{eq:rpcap}) on top of
Eqs.~(\ref{eq:rpm}) and (\ref{eq:r}) as NNNLO${}_\mathrm{approx}$+NNLL.

Power-like corrections were found to be indispensable for a realistic
description of the experimental data of 
$\langle n_h\rangle_q$, $\langle n_h\rangle_g$, and $r$
\cite{Capella:1999ms,Dokshitzer:1992iy}.
Following Refs.~\cite{Capella:1999ms,Dokshitzer:1992iy}, we include them by
multiplying $r_+$ in Eq.~(\ref{eq:rpcap}) with the factor
(to obtain it we collect all terms $\sim \gamma_0/\mu $ in the Appendix 1
of \cite{Capella:1999ms})
\begin{equation}
  1+(1+\frac{n_f}{27})\frac{\mu_{\mathrm{cr}}}{\mu}\gamma_0
+\mathcal{o}\left(\frac{\mu_{\mathrm{cr}}}{\mu}\gamma_0\right) ,
\label{eq:power}
\end{equation}
where $\mu_\mathrm{cr}$ is a critical scale parameter to be fitted.
In the MLLA approach, $\mu_\mathrm{cr}=K_\mathrm{cr}\Lambda_\mathrm{QCD}$
usually serves as the initial point of the evolution, which is implemented with
the basic variables $Y=\ln(\mu/\mu_0)$ and $\lambda=\ln K_\mathrm{cr}$.
The most frequent choice, $\lambda=0$, corresponds to the
{\it limiting-spectrum} approximation \cite{Azimov:1984np}.
Other recent choices include $\lambda=1.4$ and $\lambda=2.0$
\cite{Perez-Ramos:2013eba}.
Since logarithmic and powerlike corrections become comparable at small values
of $\mu^2$, a judicious choice of $\mu$ is important to prevent strong
correlations.
Motivated by Refs.~\cite{Brodsky:1976mg,Aversa:1990uv,Aguilar:2014tka}, we
choose $\mu^2=R^2Q^2+4M_\mathrm{eff}^2$, where $R$ is the jet radius,
$Q^2=\sqrt{s}$, and $M_\mathrm{eff}$ is the effective gluon mass.
We adopt $R=0.3$ as a typical value from Ref.~\cite{Aversa:1990uv} and
$M_\mathrm{eff}(Q^2)=m^2/[1+(Q^2/M^2)^\gamma]$ with $m=0.375$~GeV,
$M=0.557$~GeV, and $\gamma=1.06$ from Ref.~\cite{Aguilar:2014tka}.

\begin{figure}
  \centering
\includegraphics[scale=0.6,clip]{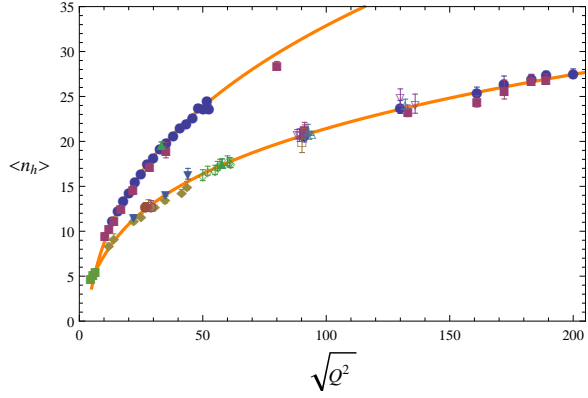}
\caption{Comparison of the experimental data of $\langle n_h(\mu^2)\rangle_q$
(lower curves) and $\langle n_h(\mu^2)\rangle_g$ (upper curves) with the
$\mathrm{NNNLO}_\mathrm{approx}+\mathrm{NNLL}$ fit to them.}
\label{fig:n}
\vskip -0.3cm
\end{figure}

We are now in a position to perform a global fit to the available measurements
of $\langle n_h\rangle_q$ and $\langle n_h\rangle_g$ for changed hadrons $h$ in
$e^+e^-$ annihilation, which were carefully compiled in
Ref.~\cite{Bolzoni:2013rsa}.
As in Ref.~\cite{Bolzoni:2013rsa}, we choose the reference
scale to be $Q_0=50$~GeV, which roughly corresponds to the geometric mean of
the smallest and largest of the occurring $\sqrt{s}$ values, and put $n_f=5$
throughout our analysis.
As may be seen in Fig.~\ref{fig:n}, our
$\mathrm{NNNLO}_\mathrm{approx}+\mathrm{NNLL}$ fit yields an excellent
description of the experimental data included in it, with a $\chi^2$ per degree
of freedom of $\chi_\mathrm{dof}^2=1.32$.
The fit parameters are determined to be
$\langle n_h(Q_0^2)\rangle_q=16.38\pm 0.05$,
$\langle n_h(Q_0^2)\rangle_g=23.87\pm 0.07$,
$K_\mathrm{cr}=7.09\genfrac{}{}{0pt}{}{+1.71}{-1.21}$, and
$\alpha_s^{(5)}(M_Z^2)=0.1205\genfrac{}{}{0pt}{}{+0.0026}{-0.0037}$,
where the errors refer to the 90\% confidence level (CL) and are evaluated as
explained in Ref.~\cite{Bolzoni:2013rsa}.
At 68\% CL, we have
   \begin{equation}
\alpha_s^{(5)}(M_Z^2)=0.1205\genfrac{}{}{0pt}{}{+0.0016}{-0.0020},
\label{eq:as}
\end{equation}
which nicely agrees with the present world average,
$\alpha_s^{(5)}(M_Z^2)=0.1181\pm0.0011$ \cite{Olive:2016xmw}.
Our fit results turn out to be very insensitive to the precise choice of $Q_0$.
The power corrections turn out to be sizeable, with
$\lambda=1.96\genfrac{}{}{0pt}{}{+0.21}{-0.19}$, 
in agreement with Ref.~\cite{Perez-Ramos:2013eba}.

\begin{figure}
\centering
  \includegraphics[scale=0.6,clip]{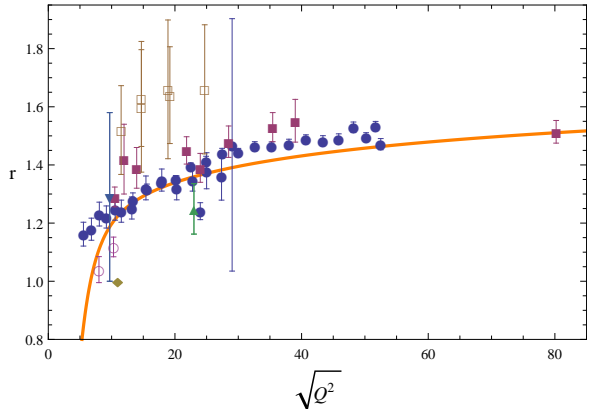}
\caption{Comparison of our $\mathrm{NNNLO}_\mathrm{approx}+\mathrm{NNLL}$
prediction of $r(\mu^2)$ with experimental data excluded from the fit.}
\label{fig:r}
\vskip -0.7cm
\end{figure}
In Fig.~\ref{fig:r}, we compare our
$\mathrm{NNNLO}_\mathrm{approx}+\mathrm{NNLL}$ prediction for $r$ with the
experimental data compiled in Ref.~\cite{Bolzoni:2013rsa}, which did not enter
our fit.
Someone can see that our results lie something below of the most of points
  that shows about a disagreements between the experimental data for
  multiplicities and their ratio $r$.

  In summary, we shown
  an unexpected, SUSY-like relationship \cite{Kniehl:2017fix} between the
NNLL-resummed first Mellin moments of the timelike DGLAP splitting functions
in real QCD, Eq.~(\ref{Basic}), which is $n_f$ independent.
Also incorporating the appropriately transformed $\mathcal{O}(\gamma_0^2)$ and
$\mathcal{O}(\gamma_0^3)$ corrections to $r_+$ as well as power-like
corrections, we shown also
(performed in  \cite{Kniehl:2017fix})
a global fit to the world data of charged-hadron
multiplicities in quark and gluon jets produced by $e^+e^-$ annihilation and so
extracted the competitive new value of $\alpha_s^{(5)}(M_Z^2)$ in
Eq.~(\ref{eq:as}), which nicely agrees with the present world average.
Notice that there is only
a 1$\%$ difference between the
result (\ref{eq:as}) and one in
  \cite{Bolzoni:2013rsa}, that shows an independence of the results
  for strong couplings of the order of usage of diagonalization and the
  limit $N \to 1$.

This research was supported in part 
by the Russian Foundation for Basic Research under Grant No.\ 16-02-00790-a.

\end{document}